\documentclass[aps,prb,twocolumn,superscriptaddress]{revtex4-1}
\usepackage{graphicx}
\usepackage{amssymb}
\usepackage{amsmath}
\usepackage{amsfonts}
\usepackage{braket}
\usepackage{color}

\begin{document}

\title{Magnon band structure of skyrmion crystals and stereographic projection approach  
} 
\author{V.~E. Timofeev}
\email{vetimofeev@etu.ru}
\affiliation{NRC ``Kurchatov Institute", Petersburg Nuclear Physics Institute, Gatchina
188300, Russia}
\affiliation{St. Petersburg Electrotechnical University ``LETI'', 197376 St. Petersburg, Russia}
\author{D.~N. Aristov}
\affiliation{NRC ``Kurchatov Institute", Petersburg Nuclear Physics Institute, Gatchina
188300, Russia}
\affiliation{St.Petersburg State University, 7/9 Universitetskaya nab., 199034
St.~Petersburg, Russia} 
 
\begin{abstract}

Using semiclassical method combined with  stereographic projection approach,   
we investigate the magnetic dynamics of the skyrmion crystal (SkX), formed in planar ferromagnet with both Dzyaloshinskii-Moriya interaction and  uniform magnetic field. The topologically non-trivial ground state of SkX is  described in stereographic projection by the complex valued  function with simple poles at skyrmions' positions.  
We use the earlier proposed ansatz for this ground state function  in the form of the sum of individual skyrmions. 
The  dynamics follows from the  second  variation of the classical action. 
Numerical analysis yields the magnon band structure of tight-binding form in accordance with previously known results.  There are two sets of bands, one set with a flat dispersion, topologically trivial and rapidly evolving with magnetic field. Another set is robust to magnetic field, characterized by pronounced dispersion and with the Berry curvature which may be sign-reversal in the Brillouin zone.  
The developed theory can be straightforwardly generalized for the analysis of magnetic dynamics in topological spin structures of other types.  

\end{abstract}

\maketitle

\section{Introduction}

Magnetic skyrmions are one of the most popular examples of   topologically protected configuration of local magnetization \cite{everschor2018perspective, roadmap2020, GOBEL20211}. The topological protection and  
relatively small size of skyrmions make them promising candidates for development  of new types of magnetic random access memory \cite{Koshibae_2015} and programmable logic\cite{Yan2021} devices.

The theoretical concept of magnetic skyrmions appeared in the work by  Belavin and Polyakov (BP) \cite{Belavin1975}, where skyrmions were found as  exact solutions of the continuous model of two-dimensional ferromagnet at $T=0$. It was also shown there that these solutions with non-zero topological charge are metastable. It was found later that Dzyaloshinskii-Moriya (DM) interaction and external magnetic field can stabilize skyrmions\cite{bogdanov1989thermodynamically, BOGDANOV1994255}, and they tend to form the skyrmion crystal (SkX).

The experimental observations of SkX were reported more than a decade ago by means of neutron scattering  \cite{muhlbauer2009skyrmion, adams11} and Lorentz transmission electron microscopy\cite{Yu2010b}. Nowadays there is a large number of compounds and systems, where different types of skyrmionic lattices with different crystal symmetries were confirmed, see the review \cite{tokura2021}.

The dynamics of skyrmionic systems is both interesting and difficult question,  still under active investigation. A single skyrmion can be viewed as a circular domain wall of small size, i.e. a magnetic bubble. One can then use the Thiele equation\cite{thiele1973} for the description of single skyrmion motion\cite{guslienko2016gyrotropic}. At the same time, the skyrmion dynamics is not limited by the motion of its center. Full equations of motion for a single skyrmion should also take into account internal excitations \cite{schutte2014magnon,lin2014internal, kravchuk2018spin}. These excitations correspond to distortions of the form of skyrmions: dilatation, elliptical deformation etc.

Just as the physics of solids is not reduced to the physics of the atoms in the lattice, so too the SkX dynamics is not reduced to the motion of individual skyrmions. It was shown that the low energy dynamics of SkX   includes the  Goldstone mode\cite{petrova2011} associated with displacement of SkX as a whole. Three types of SkX excitations,  breathing, clockwise and counterclockwise modes, have been derived numerically on a lattice \cite{mochizuki2012}. Another interesting way to describe the  low energy SkX dynamics  in terms of phason excitations was suggested in  \cite{Tatara2014}.

The spatial periodicity of SkX configuration translates into the notion of the Brillouin zone (BZ) in reciprocal space, which in turn poses the question of dispersion of SkX excitations. This question was addressed numerically  on a discrete lattice  for atomic scale N\'eel-type skyrmions in \cite{roldan2016}.  
Another calculation in the framework of linear sigma model  was done in\cite{garst2017collective}  for some realistic compounds, taking into account also dipole-dipole interaction. The dependence of the band structure on external magnetic field was studied in \cite{Diaz2020}.

One of the most interesting features of the magnon band structure in SkX is its nontrivial topological properties. It was demonstrated \cite{roldan2016,garst2017collective,Diaz2020} that certain low-energy bands have non-zero Chern numbers. It was further shown that this property results in the appearance of magnon edge states\cite{roldan2016,Diaz2020}, that could be useful for magnonics. Besides, nontrivial Berry curvature and Chern numbers should lead to thermal Hall effect \cite{Katsura2010,Matsumoto2011}.

In this paper we combine the semiclassical method and the stereographic projection approach for the investigation of SkX dynamics. Such combination is primarily useful at low temperatures, when the local magnetization saturates to its maximum value. The stereographic projection describes the direction of the equilibrium magnetization in terms of complex-valued function, $f$. The magnetic energy acquires then the form of highly non-linear functional, whose extremum is well reached by  ansatz borrowed  from the original  BP paper. \cite{Belavin1975}  Having obtained a suitable description of the classical configuration of SkX, we find the dynamics from the second variation of the classical action. \cite{Rajaraman}   The proposed semiclassical analysis leads to equations of motion structurally identical to those found previously in linear spin-wave theory formalism. \cite{schutte2014magnon} Our numerical analysis of these equations yield the dispersion laws of low-lying excitations in good agreement with previous findings.   \cite{roldan2016,garst2017collective,Diaz2020} The subsequent analysis of topological properties of the low-lying bands  by link-variable method\cite{fukui2005chern} provides a  good accuracy in evaluation of the Berry curvature and the Chern numbers.  

Among the advantages of the proposed method we can name a few. One is the possibility to work consistently in the continuum model, without digressing to lattice formulation in order to find the equilibrium magnetization configuration. 
Another is the principal possibility to investigate the dynamics of irregular skyrmion structures, e.g. liquids, which is   described by the same equations of motion with a different stereographic projection function. 
The third one is better control over the so-called zero mode, describing the motion of skyrmions system as a whole. 
We plan to explore the two latter features in subsequent publications. 

The rest of the paper is organized as follows. 
In Section \ref{sec:FitstSteps} we describe our model, the general formalism of semiclassical approach and stereographic projection, and derive the equations of motion. The discussion of several important technical issues is found here as well.   In Section \ref{sec:StaticConfig} we briefly describe the construction of static SkX configuration in stereographic projection   by the ansatz proposed earlier in \cite{Timofeev2019,timofeev2021}.  Here we also discuss the way to calculate corrections to our trial ground state, and provide corresponding formulas.  
The magnon dispersion of SkX is calculated in Section \ref{sec:BandStruct} for a number of values of external magnetic field in our model, the analysis of band dispersion in terms of tight-binding approximation is made here.  In Section \ref{sec:BandsTopology} we calculate the Berry curvature and Chern numbers for low-lying magnon bands and qualitatively discuss the results. Our final remarks and conclusions are presented in Section \ref{sec:conclusion}. Further technical details are discussed in Appendices \ref{app:Kinetic} and \ref{sec:regular}.

\section{model and general formalism} \label{sec:FitstSteps}

\subsection{Energy and Lagrangian}

We consider a minimal model of the two-dimensional (2D) non-centrosymmetric ferromagnet possessing skyrmions, with the energy of the system $E = \int d^2\mathbf{r}\, \mathcal{E}$ and  the energy density
\begin{equation}
\mathcal{E} =   \frac{C}{2}  \partial_{\mu}S^{i}\partial_{\mu}S^{i} - 
D\epsilon_{\mu ij} S^{i}\partial_{\mu}S^{j}  + B(1 - S^{3}),
\label{classicalenergy}
\end{equation}
here $C$ and $D$ are exchange and DM constants, $B$ is external magnetic field perpendicular to the plane;  $\mu=1,2$ and $i=1,2,3$; totally antisymmetric tensor  $\epsilon_{\mu ij}$  and summation over the repeated indices is assumed. This simple form of DM interaction implies cubic crystals, and in general one considers DM interaction via Lifshitz invariants  depending on the particular symmetry of the system. For zero temperature considered here we have a constraint $S^{i}S^{i}=1$ for local magnetization $\mathbf{S}(\mathbf{r})$. We assume $D,B>0$ and added unity in the last term of \eqref{classicalenergy} in order to have zero energy for the anticipated uniform ground state, $S^3=1$, at $D=0$. 

The model \eqref{classicalenergy} has two characteristic length scales, $C/D$ and $\sqrt{C/B}$. Taking  values of $C$ and $D$ as basic parameters of the material, we measure length in units of  $l=C/D$. The energy density, measured in units  of $Cl^{-2}=D^2/C$,    
can be reduced to the dimensionless form~:
\begin{equation}
\mathcal{E} =\frac{1}{2}  \partial_{\mu}S^{i}\partial_{\mu}S^{i} - 
\epsilon_{\mu ij} S^{i}\partial_{\mu}S^{j}  + b(1 - S^{3}),
\label{classenS}
\end{equation}
with $b=BC/D^2$. 

The dynamics of the magnetization without damping effects is usually described by the Landau-Lifshitz (LL) equation:
\begin{equation}
\dot{\mathbf{S}}=-\gamma_0 \, \mathbf{S}\times\mathbf{H},
\label{LLeq}
\end{equation}
where $\gamma_0$ is a gyromagnetic ratio and $\mathbf{H}=\delta {E}/\delta\mathbf{S}$ is an effective magnetic field. The LL equation \eqref{LLeq} is a convenient tool to numerical calculation and micromagnetic simulations, but  we prefer below  the analysis based on the Lagrangian formalism. The Lagrangian density $L=\int d^2\mathbf{r}\, (\mathcal{T}-\mathcal{E})$ includes the  
 kinetic  term:
\begin{equation}
\mathcal{T}=
\frac \hbar{a_0^2}(1-\cos{\theta})\dot{\varphi}\,,
\label{eq:kinetic}
\end{equation}
where  $\varphi$ and $\theta$ define the  magnetization direction $\mathbf{S}=(\cos\varphi\sin\theta,\sin\varphi\sin\theta,\cos\theta)$ and  
$a_0$ is the distance between   localized magnetic moments, arising here after passing to continuum model.

\subsection{Stereographic projection approach}

We parametrize  the magnetization in the stereographic projection approach as 
\begin{equation}
S^1 + iS^2  = \frac{2f}{1 + f\bar{f}}\,,\quad 
S^3 = \frac{1 - f\bar{f}}{1 + f\bar{f}},
\label{eq:stereo}
\end{equation}
where  the complex-valued function $f = \tan( \theta / 2)\,  e^{i \varphi}$  and its complex conjugate, $\bar{f}$,
depend in our 2D case on $z=x + i y$ and $\bar{z}=x - iy$, where $x$ and $y$ are spatial coordinates.
We note in passing that in the alternative representation  \cite{schutte2014magnon} of \eqref{eq:kinetic}, $\mathcal{T}=  - \frac \hbar{a_0^2}  {\mathcal{A}} \cdot \dot{\mathbf{S}}$,  the gauge field $  \mathcal{A}$ coincides up to a prefactor with $f$, namely 
$\mathcal{A}_1 + i \mathcal{A}_2 \propto {i f}$,   $\mathcal{A}_3 =0$.

The representation \eqref{eq:stereo} is constraint-free, $|f| \in [0,\infty)$.  
In the following we also use another field variable  $\psi = \sin (\theta/2) e^{i\varphi} $, whose Jacobian of transformation from the initial spherical angles is especially simple,  $d\mathbf{S} = \sin\theta \, d\theta \,d\varphi =2i \, df\, d\bar{f}/(1 + f\bar{f})^2 =2i\, d\psi\, d\bar {\psi} $. Notice however that, due to the inconvenient constraint $|\psi|\le 1$, we use below only the differential form, $d\psi =d {f}/(1 + f_0\bar{f}_0)$, for description of small fluctuations around the static unconstrained configuration, $f_0$. 
 
The complex function $f$, with the additional condition of homogeneity, $f\to cst $ at $r\to \infty $,
represents  a map of one two-dimensional sphere onto another $S^2 \rightarrow S^2$. Such maps can belong to different homotopy classes $\pi_{2}(S^2)=\mathbb{Z}$, with an integer number  classifying the  degree of the map called topological charge. In terms of the complex function $f(z,\bar{z})$ the topological charge is given by :
\begin{equation}
Q=\frac{1}{4\pi}\int d^2\mathbf{r} \ \frac{4(\partial_{z}\bar{f}\partial_{\bar{z} }f -\partial_{z}f\partial_{\bar{z}}\bar{f})}{(1+f \bar{f})^{2}} \,.
\label{topcharge}
\end{equation}
where $\partial_{z} = (\partial_{x}-i\partial_{y})/2$ and $\partial_{\bar{z}} = (\partial_{x}+i\partial_{y})/2$.
The kinetic part of the Lagrangian is given by 
\begin{equation}
\mathcal{T}[f]=  \frac i2 \frac{\bar f \partial_t f - f \partial _t \bar f}
{1+f \bar f}
\label{kinLagrangian}
\end{equation}
whereas the energy \eqref{classenS} is 
\begin{equation}
\begin{aligned}
\mathcal{E}& = \frac{4(\partial_{z}f\partial_{\bar{z}}\bar{f}+\partial_{z}\bar{f}\partial_{\bar{z}}f) }{(1 + f\bar{f})^{2}}  \\
&+\left\{\frac{2i (\bar{f}^{2}\partial_{\bar{z}}f + \partial_{\bar{z}}\bar{f}-\partial_{z}f - f^{2}\partial_{z}\bar{f})}{(1 + f\bar{f})^{2}} 
\right\} + \frac{2bf \bar{f}}{1 + f\bar{f}}\,,
\end{aligned}
\label{classenF}
\end{equation}
Here and below we use the curly brackets  in order to designate the terms originated from the DM term in Eq.\ \eqref{classicalenergy}. 

The state corresponding to the (local) energy minimum is found by variation of \eqref{classenF},    leading to the equation:
\begin{equation}
\begin{aligned}
{\cal D}[f] =&     2f\partial_{z}\bar{f}\partial_{\bar{z}}\bar{f}
- (1 + f \bar{f})\partial_{z}\partial_{\bar{z}}\bar{f}
\\ &  -  i \{ \bar{f}\partial_{\bar{z}}\bar{f} + f\partial_{z}\bar{f} \} + \tfrac14 b\bar{f} (1 + f\bar{f}) = 0\,, 
\end{aligned}
\label{vareq}
\end{equation}
which is time-independent. 

Generally, nonlinear equations may have multiple solutions whose search is difficult and analytically unavailable in most cases. Equation \eqref{vareq} is not an exception in this sense. We postpone the discussion of our trial function to \eqref{vareq}  until Section \ref{sec:construction}, and now discuss  a general formalism of the dynamical fluctuations around the static configuration.

\subsection{Dynamics}

In this subsection we will describe the well known formalism of semiclassical quantisation\cite{Rajaraman} and its application to our situation. 
We assume that we found the static configuration $f_0(z,\bar{z})$, 
obeying the Eq.\  \eqref{vareq}.

A simplistic way to consider the dynamics is to allow  small time-dependent fluctuations around the classical solution $f(t,z,\bar{z}) = f_0(z,\bar{z}) + \delta f (t,z,\bar{z})$. The Lagrangian then takes the form of functional Taylor series:
\begin{equation}
\mathcal{L}[f_0 + \delta f] = \mathcal{L}[f_0] + \delta f \, \mathcal{L}_1[f_0] + \tfrac12 \delta f\, \delta f\, \mathcal{L}_2 [f_0]+ \ldots
\label{eq:Taylor}
\end{equation}
where  $\mathcal{L}_n$ is the $n$th variational derivative in $f$ and dots represent higher order terms beyond the scope of our paper. The first term vanishes, $\mathcal{L}_1[f_0] =0$, for $f_0$ satisfying Eq.\ \eqref{vareq}.

The kinetic part \eqref{kinLagrangian}, in view of $\mathcal{T}[f_0]=0$,  reads 
\begin{equation}
\mathcal{T}[f_0 + \delta f]=   -\frac i2 \frac{\delta \dot{f} \,\delta \bar{f}-\delta \dot{\bar{f}}\, \delta f}{(1 + f_0 \bar{f}_0)^2}+\ldots 
\label{kinpartfull}
\end{equation}
This expression has an unconvenient denominator, which can be eliminated by  local redefinition, $\psi = \delta f/(1 + f_0\bar{f}_0)$. This choice of new variable is  corroborated by the  trivial Jacobian for $\psi$, $\bar \psi$, mentioned above.  
The kinetic part takes now a simple form:
\begin{equation}
\mathcal{T}_2= \frac 12 
\begin{pmatrix}
  \bar{\psi},& \psi
\end{pmatrix}
\begin{pmatrix}
  -i\partial_t& 0\\
  0& i\partial_t
\end{pmatrix}
\begin{pmatrix}
  \psi\\
  \bar{\psi}
\end{pmatrix}.
\label{2ndvarT}
\end{equation}
 
It is known however that the above simplistic way of treating fluctuations is ill-suited for description of the so-called zero modes. In our case this mode corresponds to the translational symmetry of the skyrmion configuration on the infinite plane, 
$f_0(\mathbf{r} ) \to f_0(\mathbf{r}-\mathbf{R})$.  
The appropriate treatment of  fluctuations including the zero mode is discussed in  \cite{Rajaraman}  and in our case is given by 
\begin{equation}
     f(\mathbf{r}) =  f_0  + (1+f_0 \bar f_0)\, \psi(\mathbf{r}-\mathbf{R}(t)) 
     \label{eq:fwithR}
\end{equation} 
here  $f_0 \equiv f_0(\mathbf{r}-\mathbf{R}(t))$ acquires the dynamics of its own. 
\cite{schutte2014magnon} 
The velocity $\partial _t{\mathbf{R}}(t)$ is also assumed to be small, although different by construction from $\psi$. We provide further details in Appendix \ref{app:Kinetic}, and set $\partial _t{\mathbf{R}}(t) \equiv 0$ from now on.

Proceeding along these conventions, we represent the 
  second order Lagrangian as 
\begin{equation}
\mathcal{L} =\frac12 
\begin{pmatrix}
  \bar{\psi},& \psi
\end{pmatrix}
\left(-i
\begin{pmatrix}
  \partial_t& 0\\
  0& -\partial_t
\end{pmatrix}
-\hat{\mathcal{H}} \right)
\begin{pmatrix}
  \psi\\
  \bar{\psi}
\end{pmatrix},
\label{Lagr2}
\end{equation}
with the Hamiltonian operator $\hat{\mathcal{H}}$ of the form 
\begin{equation}
\hat{\mathcal{H}}=
\begin{pmatrix}
  (-i\nabla + \mathbf{A})^2 + U&
   V\\
  V^*&
   (i\nabla + \mathbf{A})^2 + U
\end{pmatrix} \,.
\label{ham}
\end{equation}
Here $U$, $V$ and $\mathbf{A} = \mathbf{e}_x A_x + \mathbf{e}_y A_y$ are   functions of $f=f_0(\mathbf{r})$ and its gradients:

\begin{equation}
\begin{aligned}
U & = - 4\frac{\partial_{z}f\partial_{\bar{z}}\bar{f}+\partial_{z}\bar{f}\partial_{\bar{z}}f}{(1 + f\bar{f})^{2}} 
+ b\frac{1 - f \bar{f}}{1 + f \bar{f}}
\\
& + \left\{ \frac{2i(f^2 \partial_{z}\bar{f} + \partial_{z}f - \partial_{\bar{z}}\bar{f}- \bar{f}^2\partial_{\bar{z}}f + 2i f\bar{f})}{(1 + f \bar{f})^2}\right\} \, ,
\end{aligned}
\end{equation}
\begin{equation}
\begin{aligned}
V & = 8\frac{\partial_{z}f \partial_{\bar{z}}f(1-2 f\bar{f}) + f (1 + f \bar{f})\partial_{z}\partial_{\bar{z}}f}{(1 + f\bar{f})^{2}} \\
& -\left\{ \frac{4i(3 f^2 \partial_{z}f - \partial_{\bar{z}}f(1 - 2 f \bar{f}))}{(1 + f \bar{f})^2}\right\} - b\frac{2 f^2}{1 + f \bar{f}} \,,
\end{aligned}
\end{equation}
 \begin{equation}
\begin{aligned}
A_x & =\frac{i f \partial_x \bar{f}  -i \bar f \partial_x {f}   }{1 + f\bar{f}}   
+\left\{\frac{4 \,\mbox{Re}f}{1 + f\bar{f}} \right\},
\end{aligned}
\label{Aform}
\end{equation}
whereas $A_y$ is obtained from $A_x $ by changing $\partial_x \to \partial_y$, $\mbox{Re}f \to \mbox{Im}f$. 

The Lagrangian \eqref{Lagr2}  results in the  Euler-Lagrange equation:
\begin{equation}
-i \frac{d}{dt}
\begin{pmatrix}
  \psi\\
  \bar{\psi}
\end{pmatrix}
=\sigma_3\hat{\mathcal{H}}
\begin{pmatrix}
  \psi\\
  \bar{\psi}
\end{pmatrix},
\label{BdGeq}
\end{equation}
with $\sigma_3$ the third Pauli matrix. The energy of normal modes, $ \epsilon_n$, is found from 
\begin{equation}
\Big( \epsilon_n \, \sigma_3 - \hat{\mathcal{H}} \Big) 
\begin{pmatrix}
  u_n\\
  v_n
\end{pmatrix}  
\equiv \Big( \epsilon_n \, \sigma_3 - \hat{\mathcal{H}} \Big)  \Psi_n
=  0 \,.
\label{eq:shr}
\end{equation}

The similar  equation for excitations above the skyrmionic ground state has  previously appeared in the literature, see e.g. \cite{schutte2014magnon}. Our approach generalizes this equation to the multi-skyrmion ground state by describing it in terms of the function $f$. A particular case of  \cite{schutte2014magnon} is reproduced by using in \eqref{ham} the single skyrmion solution for $f_0$, as discussed in the next Section.

\subsection{Qualitative discussion of the Hamiltonian 
\label{sec:qualitative}}

Before proceeding further with our calculation, let us qualitatively discuss the Hamiltonian. 
The form of the Hamiltonian (\ref{ham}) resembles one of Bogoliubov-de Gennes in the theory of superconductivity, although it describes bosonic excitations instead of fermions. After standard bosonic representation of spin operators, the Hamiltonian similar to (\ref{ham}) appears in the linear spin-wave theory of antiferromagnets and, generally, of   ordered magnets with non-parallel alignment of spins. 

What differs SkX from other non-collinear magnets is the existence of the gauge field $\mathbf{A}$. 
The definition  \eqref{Aform} of $\mathbf{A}$ includes two terms, arising from exchange interaction and DM interaction.  One can easily see  that the part, stemming from DM interaction, is regular over the whole plane.  
The first term in \eqref{Aform} is more complicated and can be rewritten as:
\begin{equation}
\mathbf{A}_{exc}=\frac{i f\bar{f}}{1 + f\bar{f}} \nabla \ln{\frac{\bar{f}}{f}}\,.
\label{Aex}
\end{equation}
We show below that 
near the center of one skyrmion placed at the origin the profile function $f$ contains singular and regular parts,  
 $f_{SkX} \simeq  i /\bar z + f_{reg}$.  
 It leads to an appearance of singular part in \eqref{Aex}:
\begin{equation}
\mathbf{A}_{exc}(\mathbf{r}) = \frac{2  
}{r} \hat e_\phi + \mathbf{A}_{reg}(\mathbf{r}).
\end{equation}
This in turn leads to delta-function singularity for gauge field intensity, $\mathbf{B}  = \nabla\times \mathbf{A} $:
\begin{equation}
\mathbf{B}(\mathbf{r}) = 4 \pi \delta(\mathbf{r}) \hat{z} + \mathbf{B}_{reg}(\mathbf{r})\,.
\label{gaugeB}
\end{equation}
For a single skyrmion we have $f \to 0 $ as $r\to \infty$. The application of the Stokes' theorem gives then the zero total  flux  \cite{schutte2014magnon} 
\begin{equation}
\Phi = \int d\mathbf{r} \, \mathbf{B}(\mathbf{r}) = 4\pi + \int d\mathbf{r} \, \mathbf{B}_{reg}(\mathbf{r}) = 0.
\label{flux}
\end{equation} 

For multi-skyrmion configuration  the  field intensity $\mathbf{B}$ has a set of $\delta$-functions placed at centers of skyrmions, and this set is infinite and regular in SkX case.  One can show that the zero flux property should hold also for multi-skyrmion configuration, when the integration in \eqref{flux}  is performed over polygons, referring to individual skyrmions (presumably after Voronoi partition).  Numerically we checked that the zero flux is obtained in our approach after the integration over the unit cell in SkX case.

Let us ignore for a moment the existence of anomalous terms, i.e. set $V=0$ in \eqref{ham} 
The simplified Schr\"odinger equation reads  
\[ \left(  (-i\nabla + \mathbf{A})^2 + U - \epsilon \right) \psi =0   \]
which describes the motion of the particle in the non-uniform magnetic field and electrostatic potential, both with the same spatial periodicity.  
The $\delta$-function contributions in \eqref{gaugeB} can be removed by the unitary transformation 
\[ \psi(z) \to \psi(z) \prod_j \frac{z-z_j}{\bar z-\bar z_j} \]
with $z_j$ corresponding to centers of skyrmions, see \eqref{SkXf} below.  
The regular part of the pseudo-magnetic field in  \eqref{gaugeB} is   smooth and could be well approximated by the uniform field, with two flux quanta per unit cell. 
At first glance such commensurate field can   be ignored, following the conclusions by Azbel 
and Hofstadter. However, a closer look reveals that the derivation in \cite{Azbel1964, Hofstadter1976} started from the tight-binding form of the spectrum which implies the limit $U\to \infty$ first and then projecting the Hamiltonian onto the lowest level. In our situation all quantities, $\mathbf{A}$, $U$ and $V$, are  defined by the same $f$ and hence are of the same order,  so that  $\mathbf{A}$ cannot be eliminated.

\subsection{Orthogonality and second quantization}

For each $\Psi_n$ with the energy  $\epsilon_n$,   we observe that  the function $\bar \Psi_n \equiv \sigma_1 \Psi_n^* = \begin{pmatrix}
  v_n^*\\
  u_n^*
\end{pmatrix}  $ satisfies  Eq. \eqref{eq:shr} with $\epsilon_n \to - \epsilon_n$, thanks    to the property $\hat{\mathcal{H}} = \sigma_1\hat{\mathcal{H}}^* \sigma_1$. 

 Let us define the matrix  $\Upsilon_n = \begin{pmatrix}
 u_n,&  v_n^*\\
v_n,&  u_n^*
\end{pmatrix} $,   obeying $\hat{\mathcal{H}} \Upsilon_n = \epsilon_n \sigma_3 \Upsilon_n\sigma_3$. 
We choose the normalization $  \int d^2\mathbf{r} \,\left(   |u_n|^2 -  | v_n|^2\right) = 1$, which 
leads to  $  \int d^2\mathbf{r} \, \Upsilon_n^\dagger \hat{\mathcal{H}} \Upsilon_n =  \epsilon_n\mathbf{1}$. 
It can be further shown that the orthogonality condition holds, $  \int d^2\mathbf{r} \left( u_m^*  u_n - v_m^* v_n\right) = \delta_{mn}$, which indicates the possibility of expansion 
\begin {equation}
\begin{pmatrix}
  \psi \\
  \bar \psi 
\end{pmatrix} = \sum_n (c_n \Psi_n + c_n^* \bar \Psi_n ) 
= \sum_n \Upsilon_n
\begin{pmatrix}
  c_n   \\
   c_n^ *
\end{pmatrix}
\end{equation} 
with some complex valued coefficients, $c_n$. 
It follows that the Hamiltonian in Eq.\ \eqref{Lagr2} takes  the form 
\begin {equation}
\frac12 \int d^2\mathbf{r}
\begin{pmatrix}
  \bar{\psi},& \psi
\end{pmatrix}
 \hat{\mathcal{H}}  
\begin{pmatrix}
  \psi\\
  \bar{\psi}
\end{pmatrix}
= \frac12  \sum _n  \epsilon_n   \left( c_n^ *  c_n +  c_n    c_n^ * 
   \right) 
\end{equation} 
with the latter expression acquiring familiar form $\sum _n  \epsilon_n   ( c_n^ \dagger  c_n +  \tfrac12)$ after promoting the coefficients $c_n$ into  second quantization operators. 

We can   write the orthogonality relations in the form 
\begin {equation}
\begin{aligned} 
&
 \int d^2\mathbf{r} \, \Upsilon_n^\dagger 
 \sigma_3 \Upsilon_m\sigma_3 = \delta_{nm} 
 \,, \\ 
 &  \sum_n \Upsilon_n (\mathbf{r})
 \sigma_3 \Upsilon_n ^\dagger (\mathbf{r}')\sigma_3  = 
 \delta(\mathbf{r}- \mathbf{r}') \mathbf{1} \,,
\end{aligned}
\end{equation} 
which allows us to write the Green's function  
 \begin {equation}
\begin{aligned} 
& (\omega \sigma_3 -  \hat{\mathcal{H}}  ) G(\mathbf{r}, \mathbf{r}') = 
\delta(\mathbf{r}- \mathbf{r}') \mathbf{1} 
\, , \\    & 
G(\mathbf{r}, \mathbf{r}')  =
  \sum_n \Upsilon_n (\mathbf{r})
 ( \omega \sigma_3 - \epsilon_n ) ^{-1}\Upsilon_n ^\dagger (\mathbf{r}')   \,. 
\end{aligned}
\end{equation} 
The sums over $n$ in the formulas above are performed for all $\epsilon_n>0$.

\subsection{Zero modes}

Two modes possessing  zero energy  can be constructed as follows. The classical energy \eqref{classenF} is obviously invariant upon the (infinitesimal) shift of coordinates, $z \to z+\delta z$, $\bar z \to \bar z+\delta \bar z$, the function $f$  transforms upon this as  
$f(z,\bar{z}) \to f(z,\bar{z})  +  \delta  z\,  \partial_{ z} f +  \delta \bar z \, \partial_{\bar z} f$. 
It follows that the quadratic  form \eqref{Lagr2} vanishes for $ \Psi_{\mbox{\o}}  $, $ \bar\Psi_{\mbox\o}  $ of the form  
\begin{equation}
\begin{aligned}
       \Psi_{\mbox\o}  &= \frac1{1+f_0 \bar f_0}\begin{pmatrix}
      \partial_{\bar z} f_0 \\ \partial_{\bar z} \bar f_0
    \end{pmatrix} \,, \\
    \bar\Psi_{\mbox\o}  &= \sigma_1  \Psi_{\mbox\o} ^* = 
    \frac1{1+f_0 \bar f_0}\begin{pmatrix}
      \partial_{z} f_0 \\ \partial_{z} \bar f_0
    \end{pmatrix} \,.
\end{aligned}
\label{zeromode}
\end{equation} 
Notice that the form \eqref{zeromode} ensures the zero energy even for those $f_0$ which do not obey the extremum condition \eqref{vareq}. It means that in the latter case $ \Psi_{\mbox\o},  \bar\Psi_{\mbox\o} $ are orthogonal to the first variational derivative $\mathcal{L}_1[f_0] $, see below.

\section{Static configuration} \label{sec:StaticConfig}

\subsection{Construction of skyrmion crystal
\label{sec:construction}}

We showed previously \cite{Timofeev2019,timofeev2021} that the stereographic projection approach is an effective instrument for  description of static skyrmionic configurations including SkX. Below we sketch the way to construct the trial function $f$, adopted for SkX configuration $f_{SkX}$. 

Our starting point is a simple form of $f(\mathbf{r})$ in the  Belavin-Polyakov (BP) solution of the model of planar ferromagnet with $D=B=0$. The principal observation comes from \eqref{vareq} showing that any holomorphic or antiholomorphic function provides the local energy minimum in this case. The single skyrmion configuration of topological charge $Q=1$ is given, e.g., by $f=z_0/\bar{z}$, with $|z_0|$ being a skyrmion radius. The multiskyrmion configuration can be represented as a sum of individual stereographic functions, $f=\sum z_{j}/(\bar z - Z_j)$, with arbitrary residues, $z_j$, and positions, $Z_j$.

The inclusion of DM interaction and $b\neq0$ into the model, Eq. \eqref{classenF}, obviously complicates the problem. However, this inclusion has continuous character and leads to the smooth deformation of the BP solution, whereas  topological properties such as $Q$ should not change. Our choice  $b>0$ leads to a decrease  $f\to 0$  at the infinity, i.e. we should restore the function of the type $f=z_0/\bar{z}$ in the BP limit. The  DM interaction defines a helicity of  skyrmions (N\'eel or Bloch type), which means that $z_0$ should have a certain phase.   We propose the ansatz for a single BP-like skyrmion 
\begin{equation}
f_1=\frac{i \, z_0 \,\kappa(z\bar{z}/z_0^2)}{\bar{z}},
\label{eq:anz}
\end{equation}
where the shape of skyrmion is defined by the  real-valued function  $\kappa$ of distance, $r = \sqrt{z\bar{z}}$, from the center of  skyrmion; we fix  $\kappa(0)=1$. The {real-valued} $z_0$ gives  the size of skyrmion, and the factor $i$ in the numerator corresponds to the Bloch type of skyrmion stemming from  \eqref{classicalenergy}, \eqref{vareq}. The substitution of \eqref{eq:anz} into  \eqref{vareq}, gives the  equation for  $\kappa$ in the  form:
\begin{equation}
\kappa(b  z_0^2 \kappa^{2} - 4\kappa (z_0 + 2 \kappa') + x(b z_0^2 + 8  \kappa'^{2}))= 4x(x+\kappa^{2})\kappa''  \,,
\label{eq:kappa}
\end{equation}
with $x=z\bar{z}/z_0^2$, $\kappa'(x)=d\kappa(x)/dx$ etc. This nonlinear equation can be solved numerically for given $b$, $z_0$.

\begin{figure}[t]
\center{\includegraphics[width=0.95\linewidth]{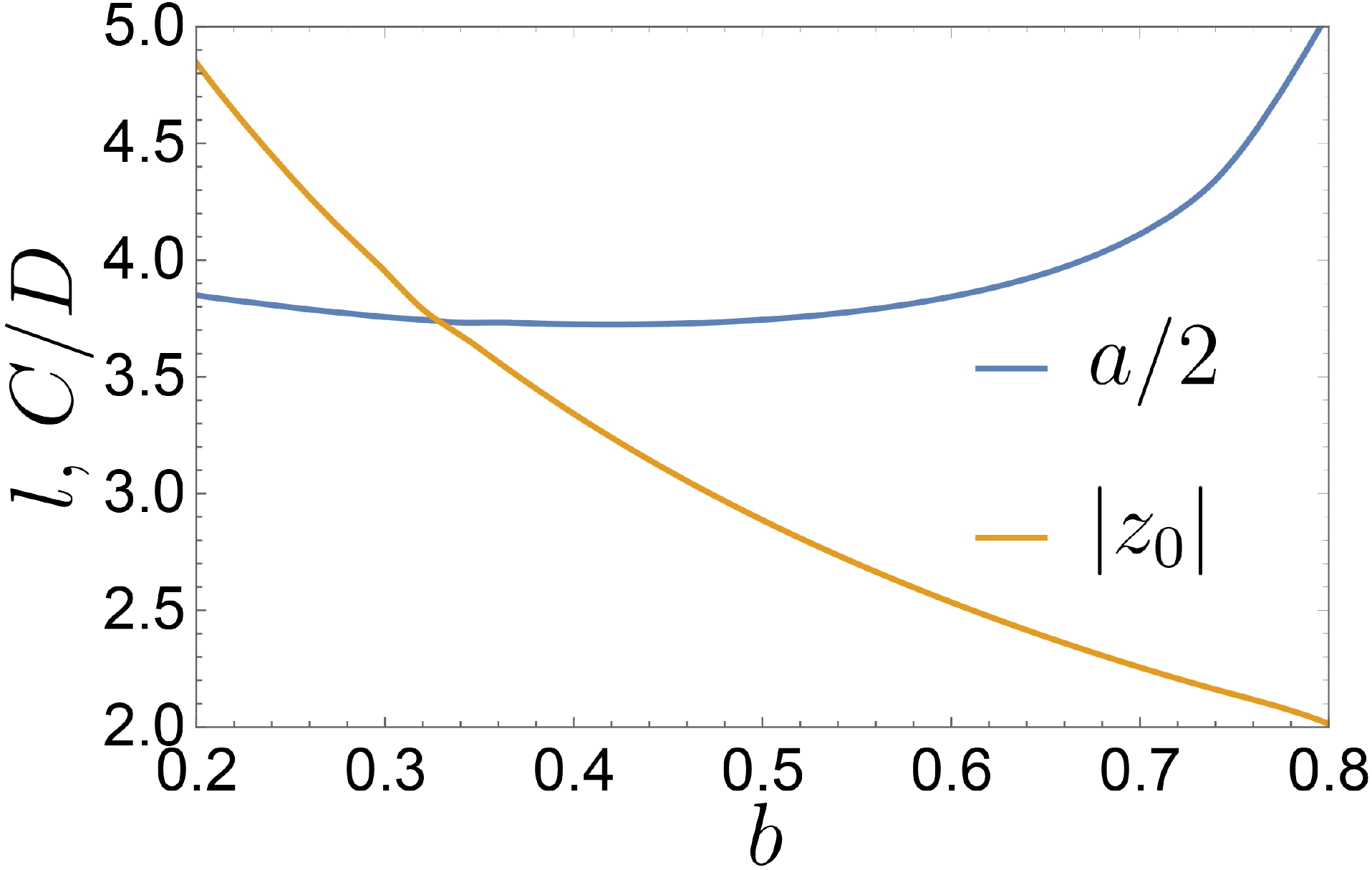}}
\caption{Optimal values of a half of the cell  parameter, $a/2$, and skyrmion radius, $|z_0|$.}
\label{fig:az0}
\end{figure}

It was shown \cite{Timofeev2019} that the ansatz (\ref{eq:anz}) allows  to take into account the interaction between skyrmions and find optimal parameters of SkX.
We model the SkX by the combination 
\begin{equation}
f_{SkX}(a,z_0) = \sum\limits_{n,m} f_1 (\mathbf{r} - n \mathbf{a}_1 - m \mathbf{a}_2),
\label{SkXf}
\end{equation}
with $\mathbf{a}_1 = (0,a)$ and $\mathbf{a}_2 = (-\sqrt{3}a/2,a/2)$, where $a$ is a cell parameter of SkX. Determining the profile of individual skyrmion for given $b$ and some  $z_0$ from Eq. \eqref{eq:kappa}, we place the skyrmions on the triangular lattice with parameter $a$. Calculating the average energy density from \eqref{classenF}, we then find  the optimal values of $z_0$ and $a$. We show the results of this calculation in Fig. \ref{fig:az0}. Interestingly, the optimal radius of skyrmions approximately follows the relation $z_0\approx 1.5/ b$ in the most relevant region, $b\in (0.3,0.7)$, which makes the profile function nearly gaussian one, $ \kappa(x) \simeq \exp(-x /2b)$.  \cite{Timofeev2019}

\begin{figure}[t]
\center{\includegraphics[width=0.95\linewidth]{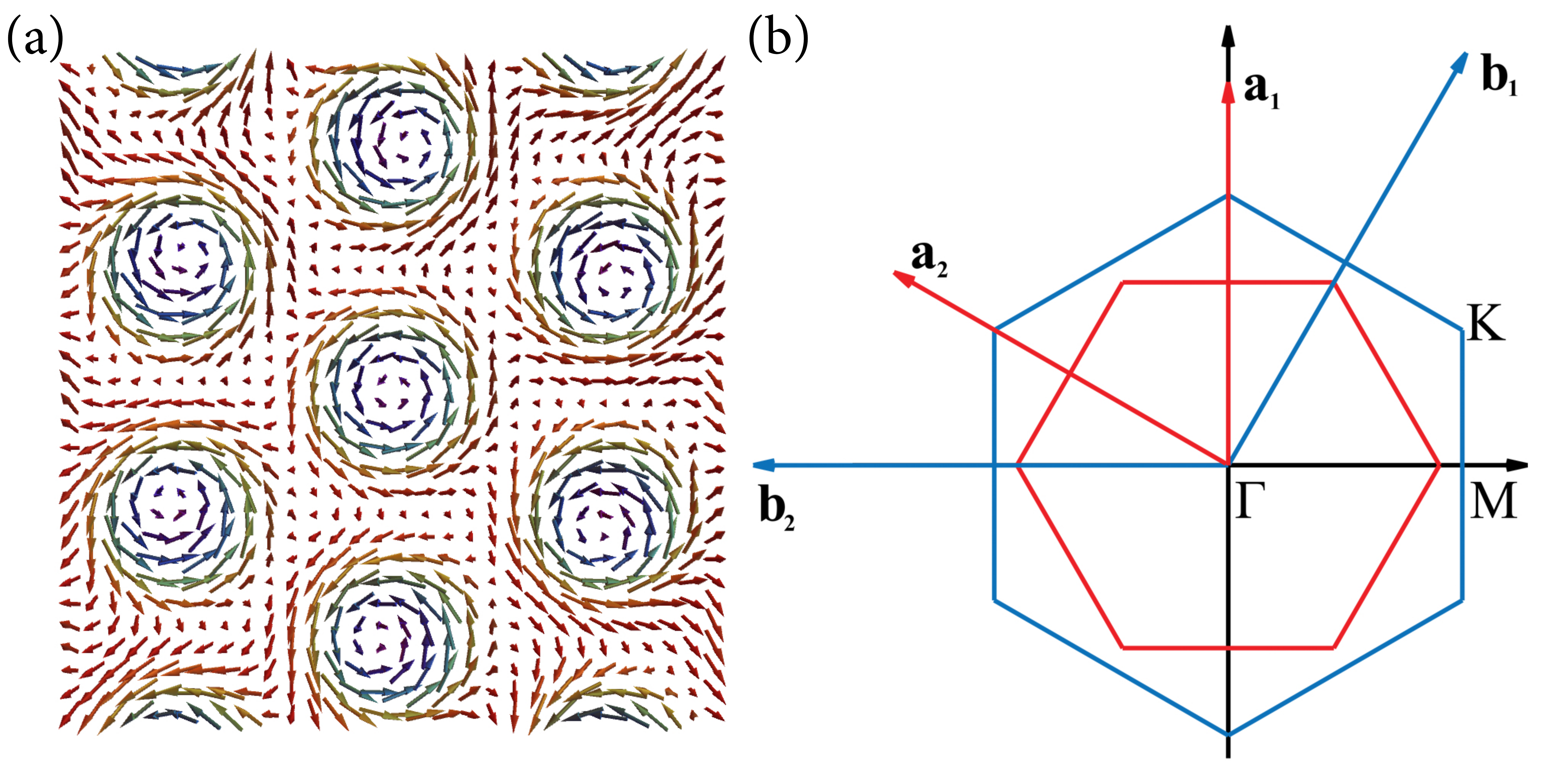}}
\caption{(a) SkX visualization (b) primitive cell and BZ.}
\label{fig:cell}
\end{figure}

The primitive vectors of the reciprocal lattice space are $\mathbf{b}_1=\frac{2\pi}{a}(1/\sqrt{3},1)$ and $\mathbf{b}_2=\frac{2\pi}{a}(-2/\sqrt{3},0)$ as shown in Fig.\ \ref{fig:cell}. Two special  points are $M = -\mathbf{b}_2 /2 $ and $ K = (\mathbf{b}_1 -\mathbf{b}_2)/3$.

\subsection{Correction to the trial ground state}
\label{sec:GScorrection}
A legitimate question arises, what is the consequence of the approximate character of our trial  ground state function \eqref{SkXf} which does not satisfy the equation \eqref{vareq}. To answer  this question, we return to 
\eqref{eq:Taylor} and represent the static part of it in the form 
\begin{equation}
\begin{aligned}
\mathcal{L}  &=  - \tfrac12 ( \begin{pmatrix}
  \bar{\psi},& \psi
\end{pmatrix}
\tilde {\mathcal{L}}_1 + H.c.) - 
 \tfrac12 
\begin{pmatrix}
  \bar{\psi},& \psi
\end{pmatrix}
\hat{\mathcal{H}} 
\begin{pmatrix}
  \psi\\
  \bar{\psi}
\end{pmatrix}, 
\\   & = 
-  \tfrac12 
\begin{pmatrix}
  \bar{\psi},& \psi
\end{pmatrix}_{new}
\hat{\mathcal{H}} 
\begin{pmatrix}
  \psi\\
  \bar{\psi}
\end{pmatrix} _{new}  + \tfrac12 
\tilde {\mathcal{L}}_1^\dagger\hat{\mathcal{H}}^{-1}  \tilde {\mathcal{L}}_1 \,, 
\end{aligned}
\end{equation}
where we use Eq.\ \eqref{vareq} and define
\begin{equation}
\begin{aligned}
\tilde {\mathcal{L}}_1 &=
\frac{8}{(1+f \bar f)^2}
\begin{pmatrix}
  {\cal D}[f]^*\\
  {\cal D}[f]
\end{pmatrix} \,,
\\ 
\begin{pmatrix}
  \psi\\
  \bar{\psi}
\end{pmatrix} _{new} 
& =\begin{pmatrix}
  \psi\\
  \bar{\psi}
\end{pmatrix}  + \hat{\mathcal{H}}^{-1}  \tilde {\mathcal{L}}_1 \,. 
\end{aligned}
\label{eq:L1def}
\end{equation}
We see that the action starts from the quadratic terms of $\begin{pmatrix}
  \bar{\psi},& \psi
\end{pmatrix}_{new}$ , which means the correction towards the true solution of \eqref{vareq} is $-\hat{\mathcal{H}}^{-1}  \tilde {\mathcal{L}}_1$. 

Next we expand  $\tilde {\mathcal{L}}_1$ in a series 
\begin{equation}
\begin{aligned}
\tilde {\mathcal{L}}_1 &= \sum_n c_n
 \Psi_n  
 \,,
\end{aligned}
\label{eq:L1expa}
\end{equation}
with the complete set of  eigenfunctions of \eqref{eq:shr} 
with $\epsilon_n>0$. The correction to the ground state function is then 
\begin{equation}
 f_{SkX} \to  f_{SkX} + \delta f =  f_{SkX} - (1 + f\bar{f}) \sum_n \frac {c_n}{ \epsilon_n}
 \psi _ n
 \end{equation}
and  the correction to the ground state energy takes  the form 
 \begin{equation}
 \delta E_0 = 
 - \tfrac12 
\tilde {\mathcal{L}}_1^\dagger\hat{\mathcal{H}}^{-1}  \tilde {\mathcal{L}}_1
= - \sum_n \frac {|c_n|^2}{ \epsilon_n} \,.
\label{energycorr2}
\end{equation}
 We use these expressions below in the discussion of our results. 
 
\section{Band structure} \label{sec:BandStruct}

\subsection{Bloch waves}
 
The spatially periodic function $f_{SkX}$ \eqref{SkXf}  is the classical static background for fluctuations, which are found in terms of normal modes of  \eqref{eq:shr}. 
The translational  symmetry of $f_{SkX}$ results in the same symmetry of $\mathbf{A},U$ and $V$ in equation \eqref{eq:shr}. 
We look for the normal modes as quasi-periodic solutions 
\begin{equation}
\Psi_{n  \mathbf{k}} = 
e^{i\mathbf{k}\mathbf{r}}\sum\limits_{\mathbf{Q}}
\begin{pmatrix}
  C^{(1)}_{\mathbf{Q}}(\mathbf{k})\\
  C^{(2)}_{\mathbf{Q}}(\mathbf{k})
\end{pmatrix}e^{i\mathbf{Q}\mathbf{r}} 
\equiv
e^{i\mathbf{k}\mathbf{r}}  \mathcal{V}_{\mathbf{k}} (\mathbf{r}) \, ,
\label{eq:blfun}
\end{equation}
where $\mathbf{k}$ lies in the first Brillouin zone,  
 index $\mathbf{Q}$ runs over the reciprocal lattice, i.e. $\mathbf{Q}=n \mathbf{b}_1 + m \mathbf{b}_2$ with $n,m$ integer. 
The sum over $\mathbf{Q}$ in \eqref{eq:blfun} corresponds to the  Bloch function, $ \mathcal{V}_{\mathbf{k}} (\mathbf{r})$,  periodic in $\mathbf{r}$-space.  

We find the Fourier transform for all potentials in Eq. \eqref{eq:shr} according to 
\begin{equation}
\mathcal{X}(\mathbf{r}) = \sum\limits_{\mathbf{Q}} \mathcal{X}_{\mathbf{Q}} e^{i\mathbf{r}\mathbf{Q}}  \,, \quad 
\mathcal{X}_{\mathbf{Q}} = \frac{1}{v}\int\limits_{cell} d^2\mathbf{r} \, \mathcal{X}(\mathbf{r}) e^{- i \mathbf{r}\mathbf{Q}},
\end{equation}
where $v= \sqrt{3} a^2 / 2$ is the unit  cell area. 
It is convenient to define a new  potential,  $\mathcal{U}=|\mathbf{A}|^{2} + U$. 
Substituting Eq.\ \eqref{eq:blfun} into Eq.\ \eqref{eq:shr} we obtain after some algebra 
\begin{widetext}
\begin{equation}
\sum\limits_{\tilde{\mathbf{Q}}}
\begin{pmatrix}
  (\mathbf{k} + \mathbf{Q})^2\delta_{\mathbf{Q}\tilde{\mathbf{Q}}} + 2\mathbf{A}_{\mathbf{Q}-\tilde{\mathbf{Q}}}(\mathbf{k}+\tilde{\mathbf{Q}}) + \mathcal{U}_{\mathbf{Q}-\tilde{\mathbf{Q}}}&
   V_{\mathbf{Q}-\tilde{\mathbf{Q}}}\\
  -V^*_{-\mathbf{Q}+\tilde{\mathbf{Q}}}&
   -(\mathbf{k} + \mathbf{Q})^2\delta_{\mathbf{Q}\tilde{\mathbf{Q}}}+ 2\mathbf{A}_{\mathbf{Q}-\tilde{\mathbf{Q}}}(\mathbf{k}+\tilde{\mathbf{Q}}) - \mathcal{U}_{\mathbf{Q}-\tilde{\mathbf{Q}}}
\end{pmatrix}
\begin{pmatrix}
  C^{(1)}_{ \tilde{\mathbf{Q}}} \\
  C^{(2)}_{\tilde{\mathbf{Q}} } 
\end{pmatrix}=
\omega_k
\begin{pmatrix}
  C^{(1)}_{\mathbf{Q}} \\
  C^{(2)}_{\mathbf{Q}} 
\end{pmatrix}.
\label{eq:matreq}
\end{equation}
\end{widetext}

After  regularization, described in Appendix \ref{sec:regular}, the coefficients $\mathcal{U}_{\mathbf{Q}}$, $\mathbf{A}_{\mathbf{Q}}$ and $ {V}_{\mathbf{Q}}$ decrease sufficiently with increasing $|\mathbf{Q}|$. It allows us to consider a finite basis in reciprocal space for (\ref{eq:matreq}). 
We choose a simple way to restrict the basis set, $\mathbf{Q}=n \mathbf{b}_1 + m \mathbf{b}_2$ with $n,m=-N, \ldots ,N$, thus  considering a large rhombus in reciprocal space. Although the symmetry of such rhombus is lower than hexagonal,  the difference becomes inessential for the large $N$ as checked by examining the obtained Bloch functions. 
 The precision of calculation depends on the basis size $N_b=2\times(2N + 1)^2$ and regularization parameter $\sigma$,   
and   exact results would correspond to   $N\rightarrow\infty$ and $\sigma\rightarrow 0$. 
The data reported below are obtained for $ N =20 $ and $\sigma =0.05$. 

 In view of relative smallness of coefficients  $ V_{ \tilde{\mathbf{Q}}}$   in \eqref{eq:matreq} the  
 upper and lower components of ``spinor'' $\begin{pmatrix}  \psi\\  \bar{\psi} \end{pmatrix}$ mix insignificantly. It means that the solutions with $\omega_k>0$ correspond largely to the upper component and vice versa. In other words,  the integral weight of coefficients   $  C^{(1)}_{\mathbf{Q}}$ is dominant for positive energies.  For this reason we choose below the  
  coefficients $  C^{(1)}_{\mathbf{Q}}$ as the most representative components of the Bloch function.

\begin{figure*}[t]
\center{\includegraphics[width=0.99\linewidth]{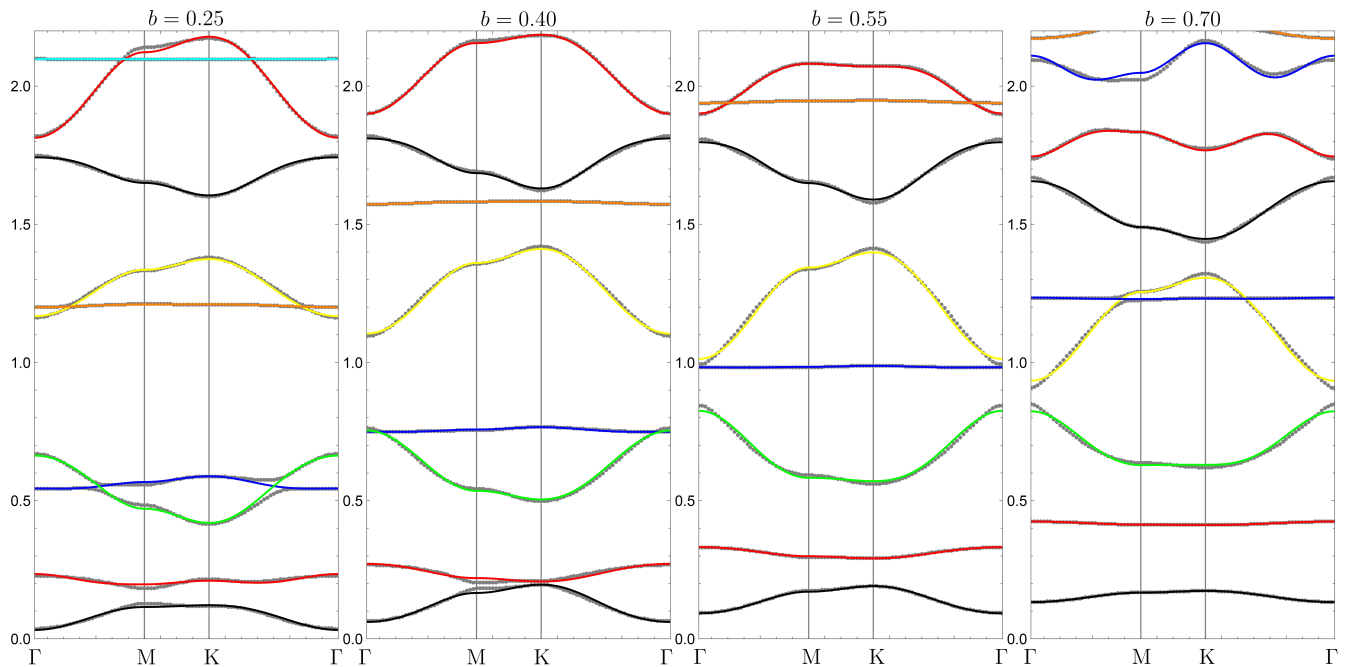}}
\caption{(Color online) Evolution of the band structure with magnetic field $b$. Gray dots show the results of numerical diagonalization of the Hamiltonian \eqref{eq:matreq} and colored curves show the fit with tight binding model \eqref{eq:tight}. Colors of branches correspond to magnetic quantum number: black ($m=2$), red ($m=3$), green ($m=0$), blue ($m=4$), yellow ($m=1$), orange ($m=5$), cyan ($m=6$), see text for additional details.}
\label{fig:b025tob07}
\end{figure*}

The results of numerical calculations for the band structure with different magnetic field $b$ is shown in Fig.\ref{fig:b025tob07}. We describe these findings in the next subsections. 

\subsection{Tight binding model}

\begin{figure}[t]
\center{\includegraphics[width=0.99\linewidth]{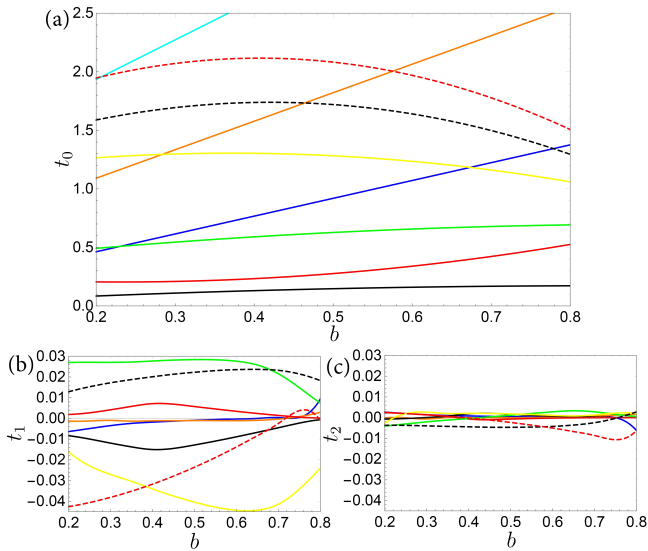}}
\caption{\label{fig:fit} 
(Color online)
Fitting parameters, $t_i$, the color of the curves  corresponds to    Fig.\ \ref{fig:b025tob07} }
\end{figure}

\begin{figure*}[t]
\center{\includegraphics[width=0.99\linewidth]{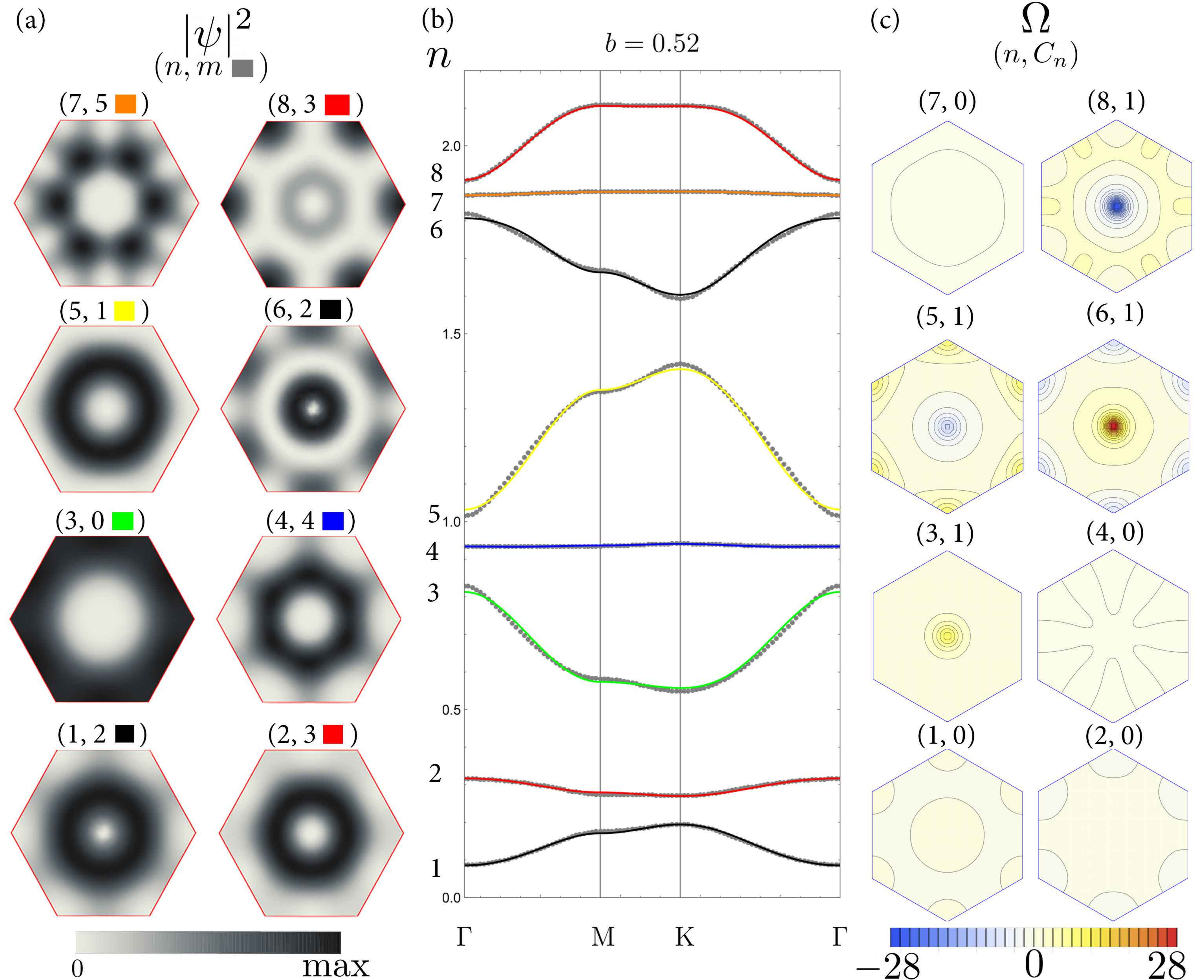}}
\caption{(Color online) The  spectrum and the wave function characteristics at   magnetic field, $b=0.52$, the value chosen by condition of non-intersection of the low-energy bands. (a) The upper spinor component density $|{\cal V}^{(1)}_{\mathbf{k}=0}(\mathbf{r})|$  in the unit cell in $\mathbf{r}$-space; the magnetic number $m$ is explained in Sec.\ \ref{sec:classification}, the color coding refers to next panel.  (b) The spectrum of eight lowest bands, gray points are the result of calculation from \eqref{eq:matreq}, color lines are the result of the fit \eqref{eq:tight}. (c) The topological properties of these bands, Berry curvature $\Omega_n$  within the BZ and Chern number  $C_n$.   }
\label{fig:052}
\end{figure*}

The low-energy band structure depicted in Fig. \ref{fig:b025tob07} and also in Fig.\ref{fig:052}b  below is remarkable, as there is no hint of parabolic dispersion expected for uniform ferromagnetic ground state, $f_0 \equiv 0$. 
At the same time one  sees  the emergence of almost flat bands.
We tried to fit each band with the tight-binding form of dispersion, including 
nearest neighbor ($t_1$) and next-to-nearest neighbor ($t_2$) hopping on triangular SkX lattice   
\begin{equation}   \label{eq:tight}    
     \begin{aligned}
    \hat{h} &=\sum_{i}t_0\, c_{i}^{\dagger}c_{i} + \sum\limits_{\langle i,j \rangle}t_1\, c_{j}^{\dagger}c_{i}+ \sum\limits_{\langle\!\langle i,j \rangle\!\rangle}t_2\, c_{j}^{\dagger}c_{i}  \,, \\ 
   &=\sum_{\mathbf{k}}(t_0 + t_1(\mathbf{k}) + t_2(\mathbf{k})) c^{\dagger}_{\mathbf{k}}c_{\mathbf{k}} \,, 
 \end{aligned} \end{equation}
  where 
  \begin{equation}     \begin{aligned}
    t_1(\mathbf{k})  &=2t_1 \left( \cos ak_y +     2 \cos \tfrac12  {ak_y}  \cos \tfrac{ \sqrt{3}} 2   ak_x   \right)
    \, , \\ 
    t_2(\mathbf{k}) &= 2t_2 \left(
    \cos  \sqrt{3} ak_x  + 
     2  \cos  \tfrac{\sqrt{3} }{2} ak_x  \cos \tfrac{3 }{2}   ak_y  \right) \,. 
    \end{aligned}
\end{equation}
We fitted the data for several low-lying modes, each with its own set of $t_j$,  and found rather satisfactory agreement. The results of this fit are shown in Fig. \ref{fig:fit} and can be summarized as follows. 

(i) All curves can largely be fitted by only two parameters, $t_0$, $t_1$. 
Setting $t_2=0$ worsens the fit insignificantly.  

(ii) The curves of ``bonding'' ($t_1 <0$) or ``antibonding'' ($t_1>0$) character evolve slowly with increasing $b$. They do not cross each other, as there is anti-crossing  property. 

(iii) The flat bands of ``nonbonding'' character ($t_1\simeq 0$) evolve rapidly with $t_0$ increasing with $b$. Upon this they pass through other bands, as can be seen in Fig. \ref{fig:fit}a (blue, orange, cyan curves). 

(iv) The finite size of our basis apparently affects two aspects of our data in Fig. \ref{fig:b025tob07}.
First, we see the anti-crossing feature, i.e.\ repulsion of non-bonding and bonding bands, which should presumably be absent. Second, the energy of the lowest band at $\Gamma$ point is finite instead of expected zero, as discussed below.

\subsection{Classification of the branches
\label{sec:classification} }

In view of the evolution of the flat bands with $b$, and their possible crossing, it is helpful to identify the bands according to the behavior of their wave-functions near the center of skyrmions.  
This identification is most easily done in terms of angular momentum, which provides the connection of SkX excitations to single skyrmion distortions.

Near the skyrmion center, $\mathbf{r}=0$ the function can be presented as $f(z,\bar{z}) \simeq e^{i\phi} F(r)$. Time-dependent fluctuations,  $f \rightarrow f + (1 + f\bar{f})\psi$, describe deformations of skyrmionic shape characterized by angular dependence of $\psi$. Expanding near the origin, $\psi(\mathbf{r}) = \sum_m \exp(im\phi)a_m(r)$, we have amplitudes $a_m(r)$ corresponding to  different partial contributions to the wave function. These coefficients can be calculated at $\Gamma$ point ($\mathbf{k}=0$) with the obtained coefficients for the Bloch function \eqref{eq:blfun} as $a_m(r) = \sum_Q exp(-im\phi_Q)J_m(Qr)   C^{(1)}_{\mathbf{Q}} $, with $\mathbf{Q} = Q (\cos \phi_Q, \sin \phi_Q)$,  $J_m(Qr)$ the Bessel function and we take     $r\ll a$ .

It turns out that for each branch of the spectrum 
one of the coefficients $a_m(r)$ for some $m$ is significantly larger  than the others. Thus we can classify the branches by the corresponding numbers $m$.   In   Fig. \ref{fig:b025tob07} and   Fig. \ref{fig:052} we mark the branches by  different colors, 
with numbers $m=0,\ldots 6$ corresponding to 
green, yellow, black, red, blue, orange and cyan, respectively. 
The modes with $m<0$ do not occur at low energies and are not shown in these Figures.

Different $m$ are related to different deformations of skyrmions. 
Consider the equation $|f(z,\bar{z})|= 1$, defining the curve on the plane, where the local magnetization lies in-plane, $S^3=0$. For solitary skyrmion such curve is a circle, whose radius is associated with the skyrmion radius. For SkX the equation $|f(z,\bar{z})|= 1$ yields a periodic structure of disconnected contours whose shape  depends on $\psi$.

The shape of the deformed skyrmion is thus given by $|f + (1 + f\bar{f}) \psi|= 1$.  Due to the form of $f$, this equation  can roughly  be presented as $A(r) + B(r) \cos(m-1)\phi=1$ with $m$ the above character of $\psi$. The mode with $m=1$ does not change the shape of skyrmion but its radius only, it is hence the breathing mode. For given $m$ the number $|m-1|$ defines  the symmetry of skyrmion deformation:  so that for $m=-1,3$ it is a second order axis,   corresponding to elliptical deformation;  $m=-2,4$ is  triangular deformation, etc. The cases of $m=2$ and $m=0$ should be considered separately.

The lowest branch  shown in Fig. \ref{fig:b025tob07} has the angular number $m=2$ and corresponds to the zero mode discussed above in Eq. \eqref{zeromode}.  We discuss it in more detail in the next subsection. 

Since the dynamics of $\psi_k$ is simply given by $\exp(i \omega_k t)$, we see that the shape of the skyrmion depends on the combination $(m-1)\phi +  \omega_k t$ with $\omega_k > 0$. It is clear then that 
the mode with $m=2$ can be associated with clockwise rotation and the mode with $m=0$ with counterclockwise rotation, in accordance with previous findings, see e.g.\  \cite{garst2017collective}.

\subsection{The lowest lying mode and correction to the ground state}
The mode $\Psi_1(\mathbf{k} =0)$ with the minimum energy,  black line ($m=2$) in Fig.\ \ref{fig:b025tob07},  attains nearly zero energy at $\Gamma$ point.  It can be compared with the analytic expression \eqref{zeromode}. Since $f_0 \sim 1/\bar z$, the behavior of  $\Psi_{\mbox\o} $ in \eqref{zeromode} is consistent with $m=2$. Normalizing $\Psi_{\mbox\o}  $ and $\Psi_1$, we find almost perfect overlap between these two modes, e.g. $\langle \Psi_{\mbox\o}  | \Psi_1(\mathbf{k} =0) \rangle = 0.993$ for $b=0.52$.   The imperfect matching is due to the fact, that the main component of the true zero mode \eqref{zeromode} is not continuous at the skyrmion positions, namely, $ \Psi_{\mbox\o}  \sim \begin{pmatrix}       z/\bar z \\ 0    \end{pmatrix} $.

Concerning the correction to the ground state due to imprecise character of our trial function, discussed in Sec. \ref{sec:GScorrection}, we expanded the first variation   \eqref{eq:L1def} in our basis, according to \eqref{eq:L1expa}. We found the small overlap of $\tilde {\mathcal{L}}_1$ only with the states with $m=1$,  particularly with the yellow branch in Fig. \ref{fig:b025tob07}. 
Using Eq. \eqref{energycorr2} we evaluate the energy density correction as 
$\delta E_0 =  7.1\cdot 10^{-4}$ per unit cell, i.e. negligibly small value. 
 

\section{Bands Topology} \label{sec:BandsTopology}

\subsection{Link-variable method}

The nontrivial topological ground state leads to the appearance of the gauge potential $\mathbf{A} $ in Eq.\eqref{ham}. As a result, the emerging band structure has unusual   properties, characterized by Berry curvature and Chern numbers of the bands. Nontrivial topology of the  bands can leads to interesting physical phenomena, such as heat Hall conductance\cite{Katsura2010}.
The topological properties of excitations in SkX were discussed previously\cite{roldan2016}
and the advantage of the stereographic projection method used here is to provide a better accuracy and details in description of these properties. 

One can calculate Berry curvature and Chern numbers, having analytical expressions for wave functions of exact band as a function of wave vector $\mathbf{k}$. In our case we numerically calculate eigenvectors of hamiltonian for any exact value of $\mathbf{k}$ and use a special method for discretized Brillouin zone  \cite{fukui2005chern}, as  described below.



We consider Bloch state
$\Psi_{n  \mathbf{k}} = e^{i\mathbf{k}\mathbf{r}}  \mathcal{V}_{n\mathbf{k}} (\mathbf{r})$ 
referring to $n$th band 
and assuming that it is a smooth function of $\mathbf{k}$. We use the expression for Berry connection 
\begin{equation}
{\cal A}_{n,\mu}(\mathbf{k}) = 
-\bra{\mathcal{V}_{n\mathbf{k}}}i\partial_{\mu}\ket{\mathcal{V}_{n\mathbf{k}}}\,, 
\end{equation}
with $\partial_{\mu} = \partial /\partial k_\mu$ and 
\[ 
\braket{\mathcal{V}_{n\mathbf{k}} |\mathcal{V}_{n\mathbf{k}'}}
= \sum\limits_{\mathbf{Q}} \left( 
  C^{(1)*}_{\mathbf{Q}}( \mathbf{k}) 
   C^{(1)}_{\mathbf{Q}}( \mathbf{k}') - 
  C^{(2)*}_{\mathbf{Q}}( \mathbf{k})
  C^{(2)}_{\mathbf{Q}}( \mathbf{k}')
  \right) \,.
\] 
The Berry curvature is 
\begin{equation}
\Omega_{n,\mu\nu}(\mathbf{k})=\partial_{\mu}{\cal A}_{n,\nu}(\mathbf{k}) - \partial_{\nu}{\cal A}_{n,\mu}(\mathbf{k})\,.
\end{equation} 
Since in 2D system the only non-trivial component $\Omega_{n,\mu\nu} $ is $\Omega_{n,12} $, we  write below simply $\Omega_{n }\equiv \Omega_{n,12}   $. The Chern number for $n$th band is obtained after the integration over BZ :
\begin{equation}
C_n = \frac{1}{2\pi}\int\limits_{\text{BZ}}\Omega_{n}(\mathbf{k}) d\mathbf{k} \,.
\end{equation} 
In numerics, 
$\mathcal{V}_{n\mathbf{k}} (\mathbf{r})$ is not analytically known and even the  smoothness of $\mathcal{V}_{n\mathbf{k}} (\mathbf{r})$ is not guaranteed. There are different approaches to calculation of $C_n$ on some mesh over BZ and we use  the method by Fukui et al. \cite{fukui2005chern}. 

To shorten the notation,  
we write $\mathcal{V}_{n\mathbf{k}} (\mathbf{r})\equiv\ket{n,\mathbf{k}}$ and consider a discrete mesh over BZ: $\mathbf{k} = i\,\delta\mathbf{k}_1 + j\,\delta\mathbf{k}_2$, with  $\delta\mathbf{k}_{\mu} = \mathbf{b}_{\mu}/N$ and indices $i$, $j$ running from $0$ to $N$. 

For small enough mesh we have an expansion 
\begin{equation}
\ket{n,\mathbf{k} + \delta\mathbf{k}_{\mu}} \approx \ket{n,\mathbf{k}} + \partial_\mu \ket{n,\mathbf{k}}\delta k_{\mu},
\end{equation}
and for normalized functions, $\braket{n,\mathbf{k}|{n,\mathbf{k} }} =1$, we can write 
\begin{equation}
\ln\braket{n,\mathbf{k}|{n,\mathbf{k} + \delta\mathbf{k}_{\mu}}} \approx \bra{n,\mathbf{k}}\partial_{\mu}\ket{n,\mathbf{k}}\delta k_{\mu} \,,
\end{equation}
so that the  Berry connection is  
\begin{equation}
{\cal A}_{n,\mu}(\mathbf{k}) \approx \frac{1}{\delta k_{\mu}} \text{Im}\ln\braket{n,\mathbf{k}|{n,\mathbf{k} + \delta\mathbf{k}_{\mu}}} \,.
\end{equation}
Since $ {\cal A}_{n,\mu}(\mathbf{k}) $  is real-valued, and  the absolute value of $\braket{n,\mathbf{k}|{n,\mathbf{k} + \delta\mathbf{k}_{\mu}}}$   can be ignored,    we   define a new \textit{link variable} as :
\begin{equation}
\begin{aligned}
U_{n,\mu}(\mathbf{k}) & =\frac{\braket{n,\mathbf{k}|{n,\mathbf{k} + \delta\mathbf{k}_{\mu}}}}{|\braket{n,\mathbf{k}|{n,\mathbf{k} + \delta\mathbf{k}_{\mu}}}|} \,,
\label{defLink}
 \\ 
{\cal A}_{n,\mu}(\mathbf{k}) & \approx -i \frac{1}{\delta k_{\mu}}\ln U_{n,\mu}(\mathbf{k})\, .
\end{aligned}
\end{equation}
 
Within the same accuracy, we have for  the Berry connection : 
\begin{equation}
\partial_{\nu}{\cal A}_{n,\mu}(\mathbf{k}) \approx - i \frac{1}{\delta k_{\nu} \delta k_{\mu}} \ln \frac{U_{n,\mu}(\mathbf{k} + \delta \mathbf{k}_{\nu})}{U_{n,\mu}(\mathbf{k})}.
\end{equation}
so that the Berry curvature can be written  as plaquette  combination 
\begin{equation}
\Omega_{n,\mu\nu}(\mathbf{k}) \approx -\frac{i}{\delta k_{\nu} \delta k_{\mu}} \ln \frac{U_{n,\nu}(\mathbf{k} + \delta \mathbf{k}_{\mu}) U_{n,\mu}(\mathbf{k})}{U_{n,\nu}(\mathbf{k}) U_{n,\mu}(\mathbf{k}+\delta\mathbf{k}_{\nu})},
\end{equation}
and when integrated over BZ, becomes the Chern number 
\begin{equation}
C_{n} \approx \frac{i}{2\pi}\sum\limits_{\text{BZ}}\ln \frac{U_{n,1}(\mathbf{k} + \delta \mathbf{k}_{2}) U_{n,2}(\mathbf{k})}{U_{n,1}(\mathbf{k}) U_{n,2}(\mathbf{k}+\delta\mathbf{k}_{1})}.
\end{equation}
We describe the results of application of this method below. 

\subsection{Results}
In order to avoid the ambiguity with identification of the intersecting bands, we chose the value of the field $b=0.52$, with the dispersion of eight lowest-lying bands shown in Fig. \ref{fig:052}b. The topological properties of these bands are shown in Fig. \ref{fig:052}c, characterized by the Berry curvature and the Chern number.  It can be seen that the flat bands, $n=4$ and $n=7$ are topologically trivial, with $\Omega_{n}(\mathbf{k}) \simeq 0$ for any $\mathbf{k}$. This is not so for the lowest bands with pronounced dispersion,  $n=1$ and $n=2$, with sizable $\Omega_{n}(\mathbf{k})$, which is however zero on the average, with the Chern numbers $C_{1}=C_{2}=0$. Here and below the integer values of   $C_{j}$ are confirmed to the accuracy $5\cdot 10^{-3}$. 
The behavior of the Berry curvature of the remaining bands, $n=3,5,6,8$, is rather interesting. It is characterized by peaks at the symmetry points, $\Gamma$ , $\mbox{K}$, $\mbox{K}'$ and nearly constant background within the whole BZ. 

The peak in $\Omega_{3}(\mathbf{k})$ at $\Gamma$ with positive amplitude provides roughly half of the integral weight, $C_3=1$, the rest coming from the smooth background. The positive peak in $\Omega_{3}(\mathbf{k})$ finds its negative counterpart of the same amplitude in $\Omega_{5}(\mathbf{k})$ at $\Gamma$ point, but the curvature of the 5th band,  $\Omega_{5}(\mathbf{k})$, is peaked with positive amplitude also at the points  $\mbox{K}$, $\mbox{K}'$ so the overall weight is again positive, $C_5=1$. 

Further, the 5th band is close in energy to the 6th band at points $\mbox{K}$ and  $\mbox{K}'$, which is manifested by the negative amplitudes $\Omega_{6}(\mathbf{k})$ there. This is however accompanied by large positive contribution  $\Omega_{6}(\mathbf{k})$ at $\Gamma$ point and  $C_6=1$ again. Finally, the large negative contribution $\Omega_{8}(\mathbf{k})$ at $\Gamma$ point is compensated by rather smooth positive  background, with $C_8=1$. 

The analysis of the bands with higher energies is more complicated, because it is hard to find the values of $b$ when  the intersection of the bands is avoided. This intersection may produce ambiguities in the definition \eqref{defLink}. 
  
Finally, we do not observe the closing and reopening of the gap between the bands, reported elsewhere for N\'eel-type skyrmions\cite{Diaz2020} for $b\simeq 0.9$.  If this phenomenon would take place in our situation, this value of the field would correspond to the instability region, when the SkX phase is less favorable in energy than  the uniform ferromagnetic state. \cite{Timofeev2019}

\section{Conclusion} \label{sec:conclusion}

We demonstrated the application of the stereographic projection method to the study of the dynamics of SkX in 2D system with DM interaction and in perpendicular magnetic field. The appearing band structure reveals a  tight-binding character of the spectrum, with two nearly non-hybridized  sets of bands, behaving differently with the increase of the magnetic field.  One set of bands  consists of flat bands, is topologically trivial, and rapidly evolves with the field. 
Another set is characterized by pronounced dispersion and non-trivial topological properties. The Berry curvature may be sizable and  sign-reversal in the Brillouin zone, with maximum values near symmetry points, whereas the integral Chern number is always non-negative for several lowest bands. 
These properties should result in existence of an edge states and unusual heat transport phenomena. 
Particularly,  possible consequences of the sign-reversal character of Berry curvature are yet to be  explored. 

The developed theory describes the dynamics of SkX in terms of  Schr\"odinger type equation for Bogoliubov spinors.   The form of the potentials in this equation is fully defined by the complex-valued function $f$ used for description of the classical ground state. In our particular study we let $f$ be the ansatz  for hexagonal ordering of Bloch-type skyrmions.  
From a more general viewpoint,  the magnetic interactions of other types will lead to different topological structures, e.g. N\'eel-type skyrmions, antiskyrmions, quadratic ordering or skyrmion liquids. In these cases  the form of the stereographic function $f$ will also be different,  however, the equation for dynamics should be mainly intact, with obvious changes taking into account   additional interactions.

\begin{acknowledgments} 
We thank  A.O. Sorokin, K.L. Metlov, A.V. Tsypilnikov  for useful discussions. The work of V.T. was supported by the Foundation for the Advancement of Theoretical Physics BASIS (grant No. 20-1-5-126-1).  
The work of D.A. was funded in part  by the Russian Foundation for Basic Research (grant No. 20-52-12019) – Deutsche Forschungsgemeinschaft (grant No. SCHM 1031/12-1) cooperation.

\end{acknowledgments}

\appendix

\section{Kinetic part of the Lagrangian
\label{app:Kinetic}}


Consider the kinetic part \eqref{kinLagrangian} and make the substitution \eqref{eq:fwithR}. We make the formal expansion in powers of $\psi$, $\bar \psi$, then we have in zeroth order : 
\begin{equation}
\mathcal{T}_0  = \tfrac 12 \mathbf{A}_0 \cdot  \dot{\mathbf{R}}
\end{equation}
with $\dot{\mathbf{R}} =  \partial _t{\mathbf{R}}(t)$ and $\mathbf{A}_0 = i ({ f \nabla \bar{f}  - \bar f \nabla {f}   })/({1 + f\bar{f}} ) 
$,  which is the first term in \eqref{Aform}. Due to properties of $\mathbf{A}_0$ the part $\mathcal{T}_0 $  disappears after integration over $\mathbf{r}$, as discussed in Sec. \ref{sec:qualitative}. The zeroth order terms may not vanish, e.g. in finite samples due to boundary effects  not discussed here, cf.\ Eq.\ (18) in \cite{metlov13vortex}. 

 The terms of the first order in $\psi$, $\bar \psi$, up to a full time derivative, have the form 
 \begin{equation}
\mathcal{T}_1  =   i\,   
 \frac {\bar \psi\dot{f}  - \psi  \dot{ \bar{f} } } {1 + f\bar{f}}
=  i\,  \dot{\mathbf{R}} \cdot 
 \frac { \psi  \nabla \bar{f}  - \bar \psi\nabla {f}   } {1 + f\bar{f}}
  \,, 
 \label{eqT1}
\end{equation}
which is reduced to scalar product of    $(\bar \psi, \psi) $ with   $\Psi_{\mbox\o}$, $\bar\Psi_{\mbox{\o}} $, Eq. \eqref{zeromode}. Due to orthogonality  of eigenfunctions, $\mathcal{T}_1$  vanishes after integration over $\mathbf{r}$ as well. 

The terms quadratic in  $\psi$, $\bar \psi$ have two contributions, with the first one given by Eq.\ \eqref{2ndvarT}. Another contribution is proportional to $  \dot{\mathbf{R}}$ and after subtracting the full time derivative is reduced to (cf. \eqref{eqT1})  
 \begin{equation}
\mathcal{T}_{2,red}  =  - i\,  \dot{\mathbf{R}} \cdot 
 \frac { \psi  \nabla \bar{f}  - \bar \psi\nabla {f}   } {1 + f\bar{f}}  
 ( \psi    \bar{f}  +\bar \psi  {f} ) \,. 
 \label{T2red}
\end{equation}
The normal terms  in the last expression   
are given by  $-2 (\mathbf{A}_0 \cdot  \dot{\mathbf{R}} )\, \bar \psi   \psi$, 
which corresponds to Eq. (32) in Ref. [\onlinecite{schutte2014magnon}]. 
The existence of anomalous terms,  $ \psi   \psi$ and $\bar \psi \bar  \psi$, makes our result \eqref{T2red} different.  Anyway, the subsequent evaluation of Eq. (75) in \cite{schutte2014magnon} did not use the particular form of this interaction vertex, by setting $\dot{\mathbf{R}}=0$.

\section{Regularization details \label{sec:regular}}

In order to numerically find the  Fourier components $\mathbf{A}(\mathbf{Q}) $ in Eq.\ \eqref{eq:matreq}, we first find 
$\mathbf{B}(\mathbf{r}) = \mbox{curl } \mathbf{A}(\mathbf{r})$ with formulas \eqref{Aform}, \eqref{SkXf}. 
The $\delta$-function part of $\mathbf{B}$ is lost upon it, and we restore it by approximate expression 
\begin{equation}
\begin{aligned}
4 \pi \delta(\mathbf{r}) \mathbf{e}_{z} \rightarrow \mathbf{B}_\sigma \equiv\frac{2}{\sigma^2} e^{-\frac{r^2}{2\sigma^2}} \mathbf{e}_{z} .
\label{eq:regular}
\end{aligned}
\end{equation}
with  $\sigma$    a control parameter. 

The Fourier transform of the regular part of $\mathbf{B}_{reg}(\mathbf{r})$ is then found and used in the relation  
  $\mathbf{A}(\mathbf{k})   = -\frac{i}{k^2}  {\mathbf{k}\times(\mathbf{B}_{reg}(\mathbf{k})+\mathbf{B}_\sigma})$. 
 We added here the term $\mathbf{B}_\sigma $, which  for one skyrmion takes the form 
\begin{equation}
\mathbf{B}_\sigma = 4\pi \mathbf{e}_{z} 
 \exp(- k^2\sigma^2/2) \,, 
\end{equation}
and where  $\mathbf{k}$ is replaced by  reciprocal lattice vectors, $ \mathbf{Q}$,  in case of SkX.  Notice that by   construction we get the transverse gauge, $\mbox{div }\mathbf{A}(\mathbf{r})=0$.




\bibliography{skyrmionbib}

\end{document}